\documentclass[12pt]{article}
\usepackage{amsfonts}
\usepackage{amssymb}
\usepackage{amscd}

\begin{document}

\title{\large{\bf ADELIC  MODEL OF HARMONIC  OSCILLATOR}}

\bigskip

\author{ {\bf Branko Dragovich} \\ Institute of Physics, P. O. Box 57 \\ 11001 Belgrade, Yugoslavia}

\date{}
\bigskip

\bigskip

\bigskip

\maketitle

\begin{abstract}
{ Adelic quantum mechanics is formulated. The corresponding model
of the harmonic oscillator is considered. The adelic harmonic
oscillator exhibits many interesting features. One of them is a
softening of the uncertainty relation.}
\end{abstract}

\bigskip

\bigskip
\noindent{\bf 1.\ \ INTRODUCTION}
\bigskip

Since 1987, $p$-adic numbers have been applied in string theory
\cite{volovich1}-\cite{vladimirov1}, quantum mechanics
\cite{volovich2}-\cite{zelenov2}, and in some other parts of
theoretical \cite{dragovich2} and mathematical
\cite{vladimirov2}-\cite{dragovich3} physics. The obtained models
are $p$-adic analogues of some standard models constructed over
real numbers. In particular, $p$-adic strings are very attractive
because of their relevance to Planck scale physics and the product
(adelic) formula for string amplitudes \cite{freund1}.

Much attention has been paid to constructing $p$-adic quantum
mechanics, which has complex-valued wave functions and $p$-adic
canonical variables. It is significant that such theory does exist
and that it allows an exact solution of the harmonic oscillator.
What is so far unclear is the connection between $p$-adic quantum
mechanics and the standard one. This is not only the problem of
$p$-adic quantum mechanics but also of other $p$-adic models.

The most natural framework which offer mathematics to unify
standard and $p$-adic models is the analysis of adeles. So,
according to adelic formula in string theory the product of
standard four-point amplitude and all $p$-adic analogues is equal
to a constant. Some aspects of an adelic approach in quantum field
theory are considered in \cite{roth}.

This article is devoted to a formulation of adelic quantum
mechanics and its illustration by the harmonic oscillator. Adelic
concepts are more fundamental than those of standard or $p$-adic
quantum mechanics. The latter (standard and $p$-adic) are building
blocks of adelic quantum theory as a whole. The problem of the
connection between $p$-adic quantum mechanics and the standard one
solves in this approach is as follows: they are independent
components of adelic quantum mechanics. Standard quantum mechanics
may be considered as an approximation of the adelic one when
$p$-adic effects can be neglected.

In Sec. 2 we present some of the main properties of $p$-adic
numbers, adeles, and their analysis. Section 3 contains a
necessary review of the harmonic oscillator in standard and
$p$-adic quantum mechanics. Adelic quantum mechanics and adelic
harmonic oscillator are presented in Sec. 4. In the last section
we discuss results obtained and make some conclusions.

\bigskip

\noindent{\bf 2. \ \ ADELES}
\bigskip

The set of rational numbers ${\mathbb Q}$ is the simplest infinite
number field. Completion of ${\mathbb Q}$ with respect to the
usual absolute value gives the field of real numbers ${\mathbb
R}$. An analogues completion with respect to the $p$-adic norms
(valuations) yields the fields of $p$-adic numbers ${\mathbb Q}_p$
($p$ = a prime number). According to the Ostrowski theorem ,
${\mathbb R} \equiv {\mathbb Q}_\infty$  and ${\mathbb Q}_p$ (for
every $p$) exhaust all number fields which can be obtained by
completions of ${\mathbb Q}$.

Recall that a series
$$
  \varepsilon\, \sum_{k =- \infty}^{+\infty} a_k \, p^k \, , \qquad
  a_k \in \{0, 1, \cdots, p-1  \}\, ,    \eqno(2.1)
$$
where $\varepsilon =\pm 1$ and $a_k = 0$ for $k \geq k_0$,
represents a real number. If  $\varepsilon = 1$ and $a_k = 0$ for
$k \leq k_0$,  the series (2.1) represents a $p$-adic number in
${\mathbb Q}_p$ . The ring of $p$-adic integers ${\mathbb Z}_p$
consists of $x = \sum_{k \geq 0} a_k \, p^k$ or, in other words,
${\mathbb Z}_p = \{x \in {\mathbb Q}_p \, : |x|_p \leq 1 \} $,
where $|x|_p $  denotes the $p$-adic norm of $x$.

On the additive group  ${\mathbb Q}_p^+$  there exists the Haar
measure $d x_p$ which is invariant under translation, i.e.,
$d(x+a)_p = d x_p \, ,\, \, a \in {\mathbb Q}_p$. Also on the
multiplicative group ${\mathbb Q}_p^\ast = {\mathbb Q}_p \setminus
\{ 0\}$ there is the Haar measure $d^\ast x_p$ invariant under
multiplication: $d^\ast (b x)_p =|b|_p \, d^\ast x_p \, , \, \,
b\in {\mathbb Q}_p^\ast $. These measures are connected by the
equality
$$
d^\ast x = \frac{1}{1 - p^{-1}}\, \frac{d x_p}{|x|_p} \, .
\eqno(2.2)
$$

An adele is an infinite sequence
$$
   a = (a_\infty , a_2 , \cdots , a_p \,, \cdots) \, , \eqno(2.3)
$$
where $a_\infty \in {\mathbb R}\, , \, a_p \in {\mathbb Q}_p $
with the restriction that all but a finite number of $a_p \in
{\mathbb Z}_p $. Let ${\mathcal A}$ be the set of all adeles,
${\mathcal A}$ a ring under componentwise addition and
componentwise multiplication, and ${\mathcal A}^+$ an additive
group with respect to addition. A multiplicative group of ideles
${\mathcal A}^\ast$ is a subset of ${\mathcal A}$ with elements $b
= (b_\infty , b_2 , \cdots , b_p \,, \cdots)$ such that $b_\infty
\neq 0$  and $b_p \neq 0$  for every $p$, and $|b_p|_p = 1$ for
all except a finite number of $p$. A principal adele (idele) is a
sequence $(r, r, \cdots , r , \cdots) \in {\mathcal A}$ , where $r
\in {\mathbb Q}\, \, \, \,\,  (r \in {\mathbb Q}^\ast  = {\mathbb
Q} \setminus \{ 0 \})$. One can define a module of the ideles,
$$
|b| = |b_\infty|_\infty \,  \prod_p \,  |b_p|_p  \eqno(2.4)
$$
which for a principal idele is
$$
|r| = |r|_\infty \,  \prod_p \,  |r|_p  = 1 .  \eqno(2.5)
$$

An additive character on ${\mathcal A}^+$ is
$$
\chi (x y) = \chi_\infty (x_\infty y_\infty)\, \prod_p \, \chi_p
(x_p y_p) = \exp (- 2 \pi i x_\infty y_\infty) \, \prod_p \exp{
 2 \pi i \{  x_p y_p\}}_p  \, ,       \eqno(2.6)
$$
where $x,\, y \in {\mathcal A}^+$ and $\{ x_p y_p \}_p$ is the
fractional part of $x_p\, y_p$. A multiplicative character on
${\mathcal A}^\ast$ can be defined as
$$
\pi (b) =  \pi_\infty (b_\infty)\, \pi_2 (b_2) \cdots \pi_p (b_p)
\cdots = |b_\infty|_\infty^s \, \prod_p |b_p|_p^s = |b|^s \, ,
\eqno(2.7)
$$
where $b$ is an idele and $s \in {\mathbb C}$ (the field of
complex numbers). It is evident that only finitely many factors in
(2.6) and (2.7)  are different from unity. One can easily see that
$\chi (r) =1 $ when $r$ is a principal adele, and $\pi (r) = 1$ if
$r$ is a principal idele.

An elementary function on the group of adeles ${\mathcal A}^+$ is
$$
\varphi (x) =\varphi_\infty (x_\infty)\, \prod_p   \varphi_p
(x_p)\, ,                  \eqno(2.8)
$$
where $x\in {\mathcal A}^+ \, , \, \, \varphi_\infty (x_\infty)
\in {\mathcal S}({\mathbb R})\, , \, \varphi_p (x_p) \in {\mathcal
S}({\mathbb Q}_p) $. Note that $ \varphi (x)$ is a complex-valued
function and must satisfy the following conditions: ({\it i})
$\varphi_\infty (x_\infty)$ is an analytic function on ${\mathbb
R}$  and for any $n\in {\mathbb N}$ the expression
$|x_\infty|_\infty^n \,\, \varphi_\infty (x_\infty) \rightarrow 0$
as  $ |x_\infty|_\infty \rightarrow \infty $ ; ({\it ii})
$\varphi_p (x_p)$ is a finite and locally constant function, i.e.
$\varphi_p$ has a compact support and $\varphi_p (x_p + y_p) =
\varphi_p (x_p)$ if $|y_p|_p\leq p^{-n} \, ,\, n=n (\varphi_p) \in
{\mathbb N}$ ; ({\it iii}) for all but a finite number of $p , \,
\,  \varphi_p (x_p) = \Omega (|x_p|_p) ,$ where
$$
\Omega (|x_p|_p) = \left\{  \begin{array}{ll}
                 1,   &   |x_p|_p \leq 1,  \\
                 0,   &   |x_p|_p > 1 .
                 \end{array}    \right.
$$
All finite linear combinations  of elementary functions $\varphi
(x)$ make a set of the Schwartz-Bruhat functions $ {\mathcal
S}({\mathcal A})$.

The Fourier transform of $\varphi (x) \in {\mathcal S}({\mathcal
A})$ is
$$
\tilde{\varphi} (y) = \int_{{\mathcal A}^+} \varphi (x)\, \chi (x
y) \, dx = \int_{- \infty}^{+ \infty} \varphi_\infty (x_\infty) \,
\, e^{- 2 \pi i x_\infty y_\infty} \, dx_\infty
$$
$$
\times \prod_p \int_{{\mathbb Q}_p} \varphi_p (x_p) \,\, e^{2\pi i
\{x_p y_p\}_p}\, \, dx_p \, ,     \eqno(2.9)
$$
where $ dx = dx_\infty \, dx_2 \cdots dx_p \cdots$ is the Haar
measure on the additive group ${\mathcal A }^+$. The Mellin
transform of $ \varphi (x) \in {\mathcal S}({\mathcal A})$ is
defined with respect to the multiplicative character $\pi (x) =
|x|^s$, i.e.,
$$
\Phi (s) = \int_{{\mathcal A}^\ast} \varphi (x)\, |x|^s \, d^\ast
x =  \int_{- \infty}^{+ \infty} \varphi_\infty (x_\infty) \, \,
|x_\infty|_\infty^{s-1} \, dx_\infty
$$
$$
\times \prod_p \int_{{\mathbb Q}_p} \varphi_p (x_p) \,\,
|x_p|_p^{s-1}\, \, \frac{ dx_p }{1 - p^{-1}}\, , \,\, \, \,
{Re}\,s
> 1 \, ,\eqno(2.10)
$$
where $ d^\ast x = d^\ast x_\infty \, d^\ast x_2 \cdots d^\ast x_p
\cdots$ is the Haar measure on the multiplicative group ${\mathcal
A }^\ast$.

The function $\Phi (s)$ can be continued analytically on the whole
field of complex numbers, except $s=0$ and $s=1$, where it has
simple poles with residue $- \varphi (0)$ and $\tilde{\varphi}
(0)$, respectively. $\Phi (s)$ satisfies the Tate formula
$$
\Phi (s) = \tilde{\Phi} (1-s)\, ,    \eqno(2.11)
$$
where $\tilde{\Phi}$ is the Mellin transform of $\,
\tilde{\varphi}$.

Let us note that any other necessary information on $p$-adic
numbers, adeles, and their analysis can be found in
\cite{schikhof,gelfand,vladimirov2}.

\bigskip

\noindent{\bf 3. \ \ REAL AND $p$-ADIC HARMONIC OSCILLATOR}
\bigskip

The harmonic oscillator is a very attractive theoretical model
because of its exact solvability and many applications. The
corresponding Hamiltonian is
$$
H = \frac{1}{2\, m}\, k^2  + \frac{m\, \omega^2}{2}\, q^2 \, ,
\eqno(3.1)
$$
where $q$ and $k$ are position and momentum, respectively. The
evolution of classical state can be presented in the form
$$
\left( \begin{array}{l} q (t) \\ k (t)  \end{array} \right) =
T_t\, \left( \begin{array}{l} q  \\ k   \end{array} \right) = T_t
\, z, \quad  T_t = \left( \begin{array}{lc} \cos{\omega t} & (m \omega)^{-1} \sin{\omega t} \\
- m \omega \sin{\omega t}  & \cos{\omega t}
\end{array} \right)\, , \eqno(3.2)
$$
where $q = q (0), \, \, k = k (0)$. In the real case $0\neq m,\,
\omega, \, q,\, k,\, t \in {\mathbb R}$ and in $p$-adic one $0\neq
m,\, \omega, \, q,\, k,\, t \in {\mathbb Q}_p$ with conditions
$|\omega t|_p \leq p^{-1}$ for $p\neq 2$ and $|\omega t|_2 \leq
2^{-2}$, which represent convergence domains for the $p$-adic
expansions of $\cos{ \omega t}$ and $\sin{ \omega t}$ (we shall
denote these domains by $G_p$). In standard (over real numbers)
quantum mechanics the harmonic oscillator is given by the
Schr\"odinger equation
$$
\frac{d^2 \psi}{dx^2} + \frac{2 m}{\hbar^2} \, \left( E - \frac{m
\omega^2}{2}\, x^2 \right)\, \psi = 0       \eqno(3.3a)
$$
or
$$
\frac{d^2 \psi}{d\xi^2} +  \left( \frac{2 E}{\hbar \omega} - \xi^2
\right)\, \psi = 0 \, ,      \eqno(3.3b)
$$
where
$$
\xi = \left( \frac{m \omega}{h} \right)^{\frac{1}{2}} \, x
\,\sqrt{2 \pi}          \eqno(3.4)
$$
is a dimensionless position coordinate. (From now on we shall use
$m = \omega = h = 1 .$)  As is well known, the physical solutions
to (3.3b) are orthonormal eigenfunctions
$$
\psi_n (x)  = \frac{2^{\frac{1}{4}}}{(2^n n!)^{\frac{1}{2}}}\,
e^{-\pi x^2}\, H_n (x \sqrt{2 \pi})\, ,  \eqno(3.5)
$$
where $H_n (x \sqrt{2 \pi}) \, \,\,\, (n = 0, 1, 2, \cdots)$ are
the Hermite polynomials. One can easily show that $\psi_n (x) \in
{\mathcal S}({\mathbb R})$.

In $p$-adic quantum mechanics, which we shall adopt here,
canonical variables are $p$-adic numbers and wave functions are
complex valued (for quantum mechanics of $p$-adic valued
functions, see \cite{volovich2,khrennikov1,khrennikov2}). Since $x
\in {\mathbb Q}_p$ and $\psi^{(p)} (x_p) \in {\mathbb C }$ the
Hamiltonian quantization procedure does not work.

According to the Vladimirov-Volovich approach
\cite{volovich2,volovich3,volovich4} $p$-adic quantum mechanics is
given by a triple $(L_2 ({\mathbb Q}_p)\, , W_p (z)\, , U_p (t))$,
where $ {\mathbb Q}_p$ is the field of $p$-adic numbers, $ z =
\left(\begin{array}{l} q  \\ k   \end{array} \right)$ is a point
of $p$-adic classical phase space, and $t$ is a $p$-adic time.
$L_2 ({\mathbb Q}_p)$  is the Hilbert space of complex-valued
square integrable functions with respect to the Haar measure on
${\mathbb Q}_p\, , \, \, W_p (z)$ is a unitary representation of
the Heisenberg-Weyl group on $L_2 ({\mathbb Q}_p)$, and $U_p (t)$
(the evolution operator) is a unitary representation on $L_2
({\mathbb Q}_p)$ of a subgroup $G_p$ of the additive group
${\mathbb Q}_p\,$.

The operator $W_p (z)$ realizes the Weyl representation of
commutation relations and has the form
$$
W_p (z)\, \psi^{(p)} (x) = \int_{{\mathbb Q}_p} W_p (z; x,y)\,
\psi^{(p)} (y)\, dy \, , \quad \psi_p \in L_2 ({\mathbb Q}_p)\,
,\eqno(3.6)
$$
with the kernel
$$
W_p (z; x, y) = \chi_p (2 k x + k q)\, \delta (x - y + q)
\eqno(3.7)
$$
and gives
$$
W_p (z)\, \psi^{(p)} (x) = \chi_p (2 k x + k q)\, \psi^{(p)} (x +
q) .                                   \eqno(3.8)
$$

The evolution operator in $p$-adic quantum mechanics is given by
$$
U_p (t)\, \psi^{(p)} (x) = \int_{{\mathbb Q}_p} {\mathcal
K}^{(p)}_t (x,y)\, \psi^{(p)} (y)\, dy \, , \eqno(3.9)
$$
where the kernel for the harmonic oscillator is
$$
{\mathcal K}^{(p)}_t (x,y)  = \lambda_p (2 t)\, \left| \frac{1}{t}
\right|_p^{\frac{1}{2}} \, \, \chi_p \Big( \frac{x y}{\sin t} -
\frac{x^2 + y^2}{2 \tan t} \Big) \, , \quad t\in G_p \setminus \{
0\} \, ,     \eqno(3.10a)
$$
$$
{\mathcal K}^{(p)}_0 (x,y) = \delta_p (x - y) \, ,
\eqno(3.10b)
$$
where $\delta_p (x - y)$ is a $p$-adic analogue of the Dirac
$\delta${-function}.

If $t\in {\mathbb Q}_p$ has the canonical expansion
$$
t = p^\nu \, (t_0 + t_1 p + t_2 p^2 + \cdots )\, , \quad \nu \in
{\mathbb Z}\, ,\, t_0\neq 0\, , \, 0\leq t_i\leq p-1 \, ,
\eqno(3.11)
$$
then the number-theoretic function $\lambda_p (t)$ is
$$
\lambda_p (t) = \left\{ \begin{array}{lll} 1 , \quad & \nu = 2 k, \quad & \\
\big( \frac{t_0}{p} \big) , \quad & \nu = 2 k +1 , \quad & p\equiv 1 (mod\, 4) ,  \\
i\, \big( \frac{t_0}{p} \big) , \quad & \nu = 2 k +1 , \quad &
p\equiv 3 (mod\, 4) ;
\end{array}                     \right.    \eqno(3.12a)
$$
$$
\lambda_2 (t) = \left\{ \begin{array}{ll} \frac{1}{\sqrt{2}}\, [1 + (-1)^{t_1}\, i] , \quad & \nu = 2 k \, ,  \\
\frac{1}{\sqrt{2}} \, (-1)^{t_1 + t_2}\, [1 + (-1)^{t_1}\, i] ,
\quad & \nu = 2 k +1 \, ,
\end{array}                     \right.    \eqno(3.12b)
$$
where $\big( \frac{t_0}{p} \big)$ is the Legendre symbol and $k\in
{\mathbb Z} $.  The analogous kernel in the real case
\cite{volovich6} is
$$
{\mathcal K}^{(\infty)}_t (x,y)  = \lambda_\infty (\sin t)\,
\left| \frac{1}{\sin t} \right|_\infty^{\frac{1}{2}} \, \, \exp{
2\pi i \Big(\frac{x^2 + y^2}{2 \tan t} - \frac{x y }{\sin t}
\Big)} \, , \eqno(3.13a)
$$
$$
{\mathcal K}^{(\infty)}_0 (x,y) = \delta_\infty (x - y) \, ,
\eqno(3.13b)
$$
where $| \, \, |_\infty$ denotes the usual absolute value and
$$
\lambda_\infty (t) = \left\{ \begin{array}{ll} \frac{1}{\sqrt{2}}\, (1 - i) , \quad & t> 0 \, ,  \\
\frac{1}{\sqrt{2}} \,  (1 + i ) , \quad &  t< 0 \, .
\end{array}                     \right.    \eqno(3.14)
$$
The operator $U_p (t)$ satisfies the relation
$$
U_p (t) \, W_p (z) \, U_p^{-1} (t)  = W_p (T_t z).    \eqno(3.15)
$$

A character $\chi_p (\alpha t)$ can be an eigenvalue of the
operator $U_p (t)$ for the harmonic oscillator if and only if
$\alpha \in I_p \subset {\mathbb Q}_p $ takes the following forms:
$$
\alpha = 0 \, ,                          \eqno(3.16a)
$$
$$
\alpha = p^{-\nu} \, (\alpha _0 +\alpha_1 p +\cdots +
\alpha_{\nu-2} p^{\nu -2}) \, ,  \alpha_0 \neq 0 \, , \, 0\leq
\alpha_i \leq p -1\, ,                 \eqno(3.16b)
$$
where $(\it i) \, \nu \geq 2$ for $p\equiv 1 (mod\, 4) \, , \, \,
(\it ii) \, \, \nu = 2 n \, \, (n\in {\mathbb N}) $ for $p \equiv
3 (mod\, 4)$, and $(\it iii ) \, \, \nu \geq 4 \, , \alpha_0
=\alpha_1 = 1$ for $p = 2$. It means that $\alpha$ has discrete
values and may be considered as a $p$-adic energy of the harmonic
oscillator.

The corresponding eigenfunctions satisfy the equation
$$
U_p (t)\, \psi_\alpha^{(p)} (x)  =\chi_p (\alpha t)\,
\psi_\alpha^{(p)} (x) .    \eqno(3.17)
$$
The value $\alpha =0$ corresponds to a vacuum state which is
invariant under $U_p (t)$, i.e.
$$
U_p (t)\, \psi_0^{(p)} (x)  = \psi_0^{(p)} (x) . \eqno(3.18)
$$

The Hilbert space $L_2 ({\mathbb Q}_p)$ can be presented as a
direct sum of mutually orthogonal subspaces, i.e.,
$$
L_2 ({\mathbb Q}_p) =\oplus_{\alpha \in I_p} \, H_\alpha^{(p)} \,
.         \eqno(3.19)
$$
The dimensions of $H_\alpha^{(p)}$ are as follows: $(\it i)$ when
$p\equiv 1 (mod \, 4)$, dim $H_\alpha^{(p)} = \infty $ for every
possible $\alpha$; $(\it ii)$ when $p \equiv 3 (mod\, 4)$, dim
$H_0^{(p)} =1$  and dim $H_\alpha^{(p)} = p+1$ for $|\alpha|_p
\geq p^{2 n} \, \, (n \in {\mathbb N})$; and $(\it iii)$ when $p =
2$, dim $H_0^{(2)}  = $ dim $H_\alpha^{(2)} = 2$ for $|\alpha|_2 =
2^3$ and dim $H_\alpha^{(2)} = 4$  for $|\alpha|_2 \geq 2^4 $. Any
dimension determines the number of linearly independent
eigenfunctions which correspond to the degenerate eigenvalue
$\chi_p (\alpha t)$. So far eigenfunctions $\psi_\alpha^{(p)} (x)$
are obtained \cite{volovich5,zelenov2} in an explicit form for the
vacuum state $\psi_0^{(p)} (x)$ and for some higher states
$(\alpha \neq 0)$. These eigenfunctions belong to ${\mathcal S}
({\mathbb Q}_p)$. From here we mainly restrict consideration to
vacuum states.

The orthonormal vacuum eigenfunctions of the $ U_p (t)$ for the
harmonic oscillator are: $(\it i)\, \, \varphi_0^{(p)} (x)  =
\Omega (|x|_p)\, , \, \, \varphi_\nu^{(p)} (x)
=p^{-\frac{\nu}{2}}\, (1 - p^{-1})^{-\frac{1}{2}}\, \chi_p{(\tau
x^2)}  \, \delta (p^\nu - |x|_p)\, , \, \, \nu \in {\mathbb N}, \,
\tau^2 = -1,$ for $p\equiv 1 (mod \, 4)\, ; \,\, (\it ii) \, \, \
\varphi_0^{(p)} (x) =\Omega (|x|_p)$ for $p\equiv 3 (mod \, 4)$;
and $(\it iii) \, \, \varphi_0^{(2)} (x) =\Omega (|x|_2) \, , \,
\varphi_1^{(2)} (x) =2 \, \Omega (2 |x|_2) - \Omega (|x|_2) $ for
$p = 2$. Here, $\delta (p^\nu -|x|_p)$ is an elementary function
defined \cite{volovich6} as
$$
\delta (p^\nu -|x|_p) =\left\{\begin{array}{ll} 1\, , \quad &
|x|_p = p^\nu \, , \\
0\,, \quad & |x|_p \neq p^\nu .
\end{array} \right.    \eqno(3.20)
$$

\bigskip

\noindent{\bf 4. \ \ ADELIC HARMONIC OSCILLATOR}
\bigskip

We shall consider adelic quantum mechanics as a triple $(L_2
({\mathcal A})\, , W (z)\, , U (t))$, where ${\mathcal A}$ is a
ring of adeles, $ z = \left(\begin{array}{l} q  \\ k   \end{array}
\right)$ is an adelic point of a classical phase space, and $t$ is
an adelic time.  $ L_2 ({\mathcal A})$ is the Hilbert space of
complex-valued square integrable functions with respect to the
Haar measure on ${\mathcal A} , \, \, W (z)  $ is a unitary
representation of the Heisenberg-Weyl group on $L_2 ({\mathcal
A})$,  and $U (t)$  (the evolution operator) is a unitary
representation on $L_2 ({\mathcal A})$ of a subgroup $G$ of the
additive group ${\mathcal A}^+$.

An orthonormal basis of the adelic Hilbert space for the harmonic
oscillator is
$$
\psi_{\alpha \beta} (x) =  \psi_{n 0}^{(\infty)} (x_\infty)
\prod_p \psi_{\alpha_p \, \beta_p}^{(p)} (x_p)\, , \eqno(4.1)
$$
where  $\psi_{n 0}^{(\infty)} (x_\infty) \equiv \psi_n (x_\infty)
$  and $ \psi_{\alpha_p \, \beta_p}^{(p)} (x_p)$ are orthonormal
eigenfunctions in real and $p$-adic cases, respectively. By
$\alpha = ( n, \alpha_2 \,, \cdots, \alpha_p \,, \cdots)$ and
$\beta = (0 , \beta_2 \,, \cdots ,\beta_p \,, \cdots)$ we denote
adelic indices, which characterize energy levels and their
degeneration. According to \cite{volovich5} and the preceding
section, for $p\geq 3$ and $|x_p|_p \leq 1$ all $p$-adic
eigenfunctions are
$$
\psi_{0 0}^{(p)} (x_p) \equiv  \varphi_0^{(p)} (x_p) = \Omega
(|x_p|_p) .                            \eqno(4.2)
$$
Thus for any value of the adelic variable $x$ one has
$$
\psi_{\alpha_p \, \beta_p}^{(p)} (x_p) \neq \Omega (|x_p|_p)
\eqno(4.3)
$$
only for a finite number of primes $p$. In other words, in (4.1)
all but a finite number of $\psi_{\alpha_p \, \beta_p}^{(p)}
(x_p)$ are vacuum states $  \varphi_0^{(p)} (x_p) = \Omega
(|x_p|_p)$, i.e., all except a finite number of $p$-adic indices
satisfy $\alpha_p = \beta_p = 0$.  Any $\psi (x) \in L_2
({\mathcal A})$ may be presented as
$$
\psi (x) = \sum C_{\alpha \beta}\, \psi_{\alpha \beta} (x) \, ,
\eqno(4.4)
$$
where $C_{\alpha \beta} = (\psi_{\alpha \beta}\,, \psi)$. It is
worth noting that all finite superpositions in (4.4) belong to the
set of the Schwartz-Bruhat functions ${\mathcal S}({\mathcal A})$.

According to (3.8) the adelic unitary operator $W (z)$ acts in the
following way:
$$
W (z)\, \psi (x) = \chi (2 k x + k q) \, \psi (x +q) \, ,
\eqno(4.5)
$$
where $\chi (2 k x + k q)$ is the additive character on adeles
(2.6) and $\psi \in L_2 ({\mathcal A})$. Since $x,\, q,\, k \in
{\mathcal A}$, there exists prime $p_n$ such that $|2 k_p \,x_p +
k_p\, q_p|_p \leq 1$ for all $p > p_n$, and an infinite product of
real and $p$-adic characters reduces to
$$
\chi (2 k x + k q) = \chi_\infty (2 k_\infty\, x_\infty + k_\infty
\, q_\infty) \prod_{p=2}^{p_n} \chi_p (2 k_p\, x_p + k_p\, q_p)
\,. \eqno(4.6)
$$
When $x,\, q,\, k$ are principal adeles  (rational points) one has
$\chi (2 k x + k q) = 1$ and
$$
W (z)\, \psi (x) = \psi (x+q) .  \eqno(4.7)
$$

The adelic evolution operator $U (t)$ can be defined by
$$
U (t)\, \psi (x) = \int_{{\mathcal A}}\, {\mathcal K}_t (x,y)\,
\psi (y) \, dy \, ,                       \eqno(4.8)
$$
where $U (t) = U_\infty (t_\infty) \, \prod_p \, U_p (t_p) \, , \,
\, t \in G \subset {\mathcal A}$  and $\psi (x) \in L_2 ({\mathcal
A})$. The kernel ${\mathcal K}_t (x,y)$ for the harmonic
oscillator is
$$
{\mathcal K}_t (x,y)  = {\mathcal K}_{t_\infty}^{(\infty)}
(x_\infty,\, y_\infty) \prod_p {\mathcal K}_{t_p}^{(p)} (x_p\,,\,
y_p)\, ,                \eqno(4.9)
$$
where  ${\mathcal K}_{t_p}^{(p)}$  and ${\mathcal
K}_{t_\infty}^{(\infty)}$  are given by (3.10) and (3.13).

By virtue of (3.15) and an analogous relation in the real case, it
follows that
$$
U (t)\, W (z)\, U^{-1} (t)  = W (T_t \, z) \, .  \eqno(4.10)
$$

Note that $\psi_{\alpha \beta} (x)$ in (4.1) are adelic
orthonormal eigenfunctions of the evolution operator $U (t)$.
Since ${\mathcal K}_{t}^{(p)} (x, y)$ depend on $t$ through $\sin
t$ and $\tan t$, the adelic time in $U (t)$ cannot be a principal
idele.

Let $\hat{\mathcal D}$ be an operator which acts in the Hilbert
space $L_2 ({\mathcal A})$. It is natural to define an expectation
(average) value of the corresponding observable ${\mathcal D}$ in
a state $ \psi (x) \in L_2 ({\mathcal A})$ as
$$
\langle {\mathcal D}\rangle = (\psi\, , \hat{\mathcal D} \, \psi)
=  (\psi^{(\infty)}\, , \hat{\mathcal D}_\infty \,
\psi^{(\infty)}) \, \prod_p (\psi^{(p)}\, , \hat{\mathcal D}_p \,
\psi^{(p)}) = \langle {\mathcal D}_\infty\rangle \, \prod_p \,
\langle {\mathcal D}_p \rangle \,. \eqno(4.11)
$$
When $\hat{\mathcal D}$ is not a unitary operator, one has to take
care of the convergence of the infinite product in (4.11). Let us
note that one can consider operators $\hat{\mathcal D}$ composed
only of a finite number of $p$-adic components different from the
identity operators, i.e.,
$$
\hat{\mathcal D} =  \hat{\mathcal D}_\infty  \, \prod_p \,
\hat{\mathcal D}_p =  \hat{\mathcal D}_\infty \, \prod_{p=2}^{p_n}
\, \hat{\mathcal D}_p \, ,             \eqno(4.12)
$$
where after some prime $p_n$ all $\hat{\mathcal D}_p  = 1$. For
any Schwartz-Bruhat function one can find a large enough $p_n$ for
which the unitary operators $U (t)$ and $W (z)$ may be effectively
presented in the form (4.12).

Now one can introduce an operator
$$
|x|^s_{(p_n)} =  |x_\infty|^s_{\infty}\, \prod_{p=2}^{p_n}   \,
|x_p|^s_{p} \, ,       \eqno(4.13)
$$
where $s \in {\mathbb C}$ and $p_n$ is an arbitrary prime. An
expectation value which corresponds to the operator (4.13) in the
simplest vacuum state
$$
\psi_{0 0} (x)  = 2^{\frac{1}{4}} \, e^{- \pi x_\infty^2} \,
\prod_p \, \Omega (|x_p|_p)     \eqno(4.14)
$$
is
$$
\langle |x|_{(p_n)}^s \rangle = ( \psi_{0 0} \,, \, |x|_{(p_n)}^s
\, \psi_{0 0}) $$
$$ = \sqrt{2} \,\, \Gamma\Big(\frac{s+1}{2}
\Big) \, (2 \pi)^{- \frac{s+1}{2}} \, \prod_{p=2}^{p_n} \frac{1-
p^{-1}}{1 - p^{-s-1}} \, , \, \, \, \, Re\, s > -1 \, .
\eqno(4.15)
$$

When $p_n \rightarrow \infty$, we have
$$
\langle |x|^s \rangle  = \lim_{p_n \rightarrow \infty} \langle
 |x|_{(p_n)}^s \rangle  = \sqrt{2} \,\, \Gamma\Big(\frac{s+1}{2}
\Big) \, (2 \pi)^{- \frac{s+1}{2}} \, \frac{\zeta (s+1)}{\zeta
(1)} = 0 \, ,                    \eqno(4.16)
$$
where $\zeta (s)$ is the Riemann zeta function.

In particular, from (4.16) we get
$$
  \langle |x| \rangle  = 0 \, .            \eqno(4.17)
$$

Of interest is also a knowledge of the mean square deviation
$\Delta {\mathcal D}$ , which is a measure of the dispersion
around  $\langle  {\mathcal D}\rangle$ ,
$$
\Delta {\mathcal D}  = [\langle (  {\mathcal D} - \langle
{\mathcal D} \rangle )^2\rangle]^{\frac{1}{2}} = (\langle
{\mathcal D}^2 \rangle -  \langle {\mathcal D} \rangle^2
)^\frac{1}{2} \, . \eqno(4.18)
$$
Using (4.18) we obtain
$$
\Delta |x|_{(p_n)} = \frac{1}{2} \, \Big( \frac{1}{\pi}\,
\prod_{p=2}^{p_n} \frac{1 -p^{-1}}{1 - p^{-3}} \Big)^{\frac{1}{2}}
\, \left[1 - \frac{2}{\pi} \, \prod_{p=2}^{p_n}
\frac{(1-p^{-1})(1-p^{-3})}{(1-p^{-2})^2} \right]^{\frac{1}{2}}  .
\eqno(4.19)
$$
By virtue of (4.16) one has
$$
\Delta |x| =\lim_{p_n \rightarrow \infty} \Delta |x|_{(p_n)} = 0
\, .  \eqno(4.20)
$$
The expectation value of the momentum in the simplest vacuum state
can be found in the following way:
$$
\langle |k|_{(p_n)}^s \rangle = \large( \tilde{\psi}_{0 0} (k) \,,
\, |k|_{(p_n)}^s \, \tilde{\psi}_{0 0} (k)\large) \, ,
\eqno(4.21)
$$
where $\tilde{\psi}_{0 0} (k)$ is the Fourier transform of
$\psi_{0 0} (x)$ (4.14). Using (2.9) we get
$$
\tilde{\psi}_{0 0} (k)  = 2^{\frac{1}{4}} \, e^{- \pi k_\infty^2}
\, \prod_p \, \Omega (|k_p|_p) \, ,    \eqno(4.22)
$$
i.e., $\tilde{\psi}_{0 0} = \psi_{0 0}$. It is clear that the
above obtained results for coordinate $x$ are also valid for the
momentum $k$. In particular, one obtains
$$
\langle |k|\rangle =\langle |x|\rangle = 0\, , \quad \Delta |k| =
\Delta |x|  = 0 \, .     \eqno(4.23)
$$

An uncertainty relation between the adelic position and momentum
coordinates reads
$$
\Delta x_{(p_n)} \, \Delta k_{(p_n)} = \frac{1}{4 \pi} \,
\prod_{p=2}^{p_n} \frac{1- p^{-1}}{1-p^{-3}} \, \left[1 -
\frac{2}{\pi} \, \prod_{p=2}^{p_n}
\frac{(1-p^{-1})(1-p^{-3})}{(1-p^{-2})^2} \right]  , \eqno(4.24)
$$
where the factor $1/(4 \pi)$ corresponds to the ordinary case.

One gets also interesting features applying the Mellin
transformation (2.10) to the vacuum state $\psi_{0 0} (x)$ (4.14),
which can be considered as the simplest elementary function
defined on adeles. It gives
$$
\Phi (s) = \sqrt{2}\,\, \Gamma \Big( \frac{s}{2}\Big) \,
\pi^{-\frac{s}{2}}  \, \zeta (s) \, .    \eqno(4.25)
$$
Since the Fourier transform $\tilde{\psi}_{0 0} = \psi_{0 0}$, one
obtains $\tilde\Phi (s) = \Phi (s)$. Replacing $\Phi $ and
$\tilde\Phi$ in the Tate formula (2.11) by (4.25), we have the
well-known functional relation for the Riemann zeta-function:
$$
\pi^{-\frac{s}{2}}\, \Gamma \Big( \frac{s}{2}\Big)  \, \zeta (s) =
\pi^{\frac{s-1}{2}}\, \Gamma \Big( \frac{1- s}{2}\Big)  \, \zeta
(1- s)   \, .                               \eqno(4.26)
$$

\bigskip

\noindent{\bf 5. \ \ CONCLUDING REMARKS}
\bigskip

The adelic harmonic oscillator exhibits some remarkable
mathematical properties. It is a simple, exact, and instructive
adelic model. The simplest vacuum state is also the simplest
elementary function, and its form is invariant under the Fourier
transformation. Consequently, the Mellin transform of this vacuum
state in the $x -$ and $k -$representation is the same function
which satisfies the Tate formula.

Some physical aspects of the adelic harmonic oscillator are very
interesting. According to (3.4) the dimensionless position
coordinate $\xi $ may be presented at the Planck scale in the form
$$
\xi = \sqrt {2 \pi} \, \frac{x}{l_0}\, ,
$$
where $l_0 = \big( \frac{m_0 \, \omega_0}{h}  \big)^{\frac{1}{2}}$
is the Planck length, $ m_0 = \big( \frac{h\, c}{2 \pi \, G}
\big)^{\frac{1}{2}} $ and $\omega_0 = \frac{2 \pi}{t_0} = 2 \pi \,
c^2 \, \big( \frac{2 \pi \, c}{h\, G} \big)^{\frac{1}{2}}$. In
fact our calculations are performed for $l_0 = 1$, and it seems
most natural to take $l_0$ as the Planck length. Thus, the results
 obtained for the adelic harmonic oscillator may be relevant to
 Planck scale physics. According to (4.23) one can measure
 distances which are smaller than the length $l_0$. Formula (4.24)
 contains a softening of the uncertainty relation. This is a
 consequence of the $p$-adic effects.

 On the basis of the above considerations, one can suppose that at
 distances close to $l_0$ there exist not only standard virtual
 particles but also $p$-adic ones.  The adelic particles can
 interact by means of any of these virtual objects. In the above
 case of the adelic harmonic oscillator, just virtual particles of
 the $p$-adic vacuums lead to the unusual results. So $p$-adic
 effects appear through an interaction of some real particles with
 a $p$-adic virtual matter.

Standard quantum mechanics can be considered as an approximation
of the adelic one when experimentally available distances are very
large with respect to $l_0$ (the Planck length). Namely, at very
large distances $(|x_\infty|_\infty \gg 1 \, , \, |x_p|_p \gg 1
)$, for some reasons $p$-adic states are not occupied and adelic
operators $|x|_{(p_n)}\, , \, |k|_{(p_n)}$ have to be taken with
$p_n = 0$, i.e., $|x|_{(0)} = |x_\infty|_\infty$ and  $|k|_{(0)} =
|k_\infty|_\infty$ . For these operators ($|x_\infty|_\infty$ and
$|k_\infty|_\infty$) calculations in standard and adelic quantum
mechanics give the same results.

\bigskip

\bigskip

This work is partially supported by the Russian Foundation for
Fundamental Research, Grant No. 93-011-147.

\end{document}